\def\rfr#1{Eq. (\ref{#1})}

\def\Rfr#1{Eq. (\ref{#1})}

\def\asec{$''$ cy$^{-1}$}

\def\asec{$''$ cy$^{-1}$}

\def\bar{\begin{eqnarray}}
\def\ear{\end{eqnarray}}

\def\eqi{\begin{equation}}
\def\eqf{\end{equation}}
\def\eqia{\begin{eqnarray}}
\def\eqfa{\end{eqnarray}}
\def\rp#1#2{{#1\over#2}}

\def\lb#1{\label{#1}}

\def\oc2{$\mathcal{O}(c^{-2})$}

\documentclass{ws-ijmpd}

\begin{document}

\markboth{Lorenzo Iorio} {Does Cassini allow to measure
relativistic orbital effects in the Saturnian system of
satellites?}

%
\catchline{}{}{}{}{}
%

\title{DOES CASSINI ALLOW TO MEASURE RELATIVISTIC ORBITAL EFFECTS IN THE SATURNIAN SYSTEM OF SATELLITES? }

\author{LORENZO IORIO}

\address{Viale Unit$\grave{a}$ di Italia 68, 70125, Bari (BA), Italy\\
lorenzo.iorio@libero.it}

\maketitle

\begin{history}
\received{Day Month Year} \revised{Day Month Year} \comby{Managing
Editor}
\end{history}

\begin{abstract}
In this paper we address the following question: do the recent
advances in the orbit determination of the major natural
satellites of Saturn obtained with the analysis of the first data
sets from the Cassini mission allow to detect the general
relativistic gravitoelectric orbital precessions of the mean
longitudes of such moons? The answer is still negative. The
present-day down-track accuracy is adequate for Mimas, Enceladus,
Thetys, Dione, Rhea and Titan and inadequate for Hyperion, Iapetus
and Phoebe. Instead, the size of the systematic errors induced by
the mismodelling in the key parameters of the Saturnian
gravitational field like the even zonal harmonics $J_{\ell}$ are
larger than the relativistic down-track shifts by about one order
of magnitude, mainly for the inner satellites like Mimas,
Enceladus, Thetys, Dione, Rhea, Titan and Hyperion. Iapetus and
Phoebe are not sensibly affected by such kind of perturbations.
Moreover, the bias due to the uncertainty in Saturn's $GM$ is
larger than the relativistic down-track effects for all such moons
up to two orders of magnitude (Phoebe). Thus, it would be
impossible to separately analyze the mean longitudes of each
satellite. Proposed linear combinations of the satellites' mean
longitudes would allow to cancel out the impact of the
mismodelling in the low-degree even zonal harmonics and $GM$, but
the combined down-track errors would be larger than the combined
relativistic signatures by a factor $10^3$.
\end{abstract}
\section{Introduction}
A satisfactorily empirical corroboration of a fundamental theory
requires that as many independent experiments as possible are
conducted by different scientists in different laboratories. Now,
general relativity is difficult to test, especially in the
weak-field and slow-motion approximation, valid, e.g., in our
Solar System, both because the relativistic effects are very small
and the competing classical signals are often quite larger. Until
now, Solar System tests of general relativity accurate to better
than 1$\%$ have been performed by many independent groups only in
the gravitational field of the Sun by checking the effects induced
by the gravitoelectric Schwarzschild\footnote{In regard to the
much smaller gravitomagnetic Lense-Thirring precessions\cite{LT18}
of the planetary orbits, they lie just at the edge of the
present-day accuracy in planetary ephemerides and have recently
been found in agreement with the latest measurements, although the
errors are still large\cite{Ior05a}. A $6\%$ test performed in the
gravitational field of Mars with the MGS spacecraft has recently
been reported in Ref.~\refcite{Ior06}.} part of the space-time
metric on the propagation of electromagnetic waves and planetary
orbital motions\cite{Wil06}. It is as if many independent
experiments aimed to measure fundamental physical effects were
conducted always in the same laboratory. Thus, it is worthwhile to
try to use different laboratories, i.e. other gravitational
fields, to perform such tests, even if their outcomes should be
less accurate than those conducted in the Sun's field. In
principle, the best candidates other than the Sun are the giant
planets of the Solar System like Jupiter and Saturn. To date, the
only investigations of this kind are due to Hiscock and
Lindblom\cite{HisLin79}, who preliminarily analyzed the
possibility of measuring the Einstein pericentre
precessions\cite{Ein15} of some of the natural satellites of
Jupiter and Saturn, and to Iorio and Lainey\cite{IorLan05} who
investigated the measurability of the Lense-Thirring precessions
in the system of the galilean satellites of Jupiter with modern
data sets\footnote{In their original papers Lense and
Thirring\cite{LT18} proposed to use some of the satellites of
Jupiter and Saturn. }.

The aim of this paper is to investigate if the recent improvements
in the ephemerides of the major Saturnian satellites (see
Table~\ref{param} for their orbital parameters) obtained from the
first Cassini data allows to detect at least the general
relativistic gravitoelectric precessions of their orbits.
\begin{table}
\tbl{Orbital parameters of the major Saturnian satellites: $a$ is
the semimajor axis, $e$ is the eccentricity and $i$ is the
inclination, referred to the local Laplace planes
(http://ssd.jpl.nasa.gov/?sat$\_$elem). }
{\begin{tabular}{@{}cccc@{}} \toprule
Satellite & $a$ (km) & $e$ & $i$ (deg)\\
\colrule
Mimas &  \hphantom{00}185540 & 0.0196 & \hphantom{00}1.572 \\
Enceladus &  \hphantom{00}238040 & 0.0047 & \hphantom{00}0.009 \\
Thetys & \hphantom{00}294670& 0.0001 & \hphantom{00}1.091 \\
Dione & \hphantom{00}377420 & 0.0022 & \hphantom{00}0.028 \\
Rhea & \hphantom{00}527070 & 0.0010 & \hphantom{00}0.331 \\
Titan & \hphantom{0}1221870 & 0.0288 & \hphantom{00}0.280 \\
Hyperion & \hphantom{0}1500880 & 0.0274 & \hphantom{00}0.630 \\
Iapetus & \hphantom{0}3560840 & 0.0283 & \hphantom{00}7.489 \\
Phoebe & 12947780 & 0.1635 & 175.986 \\
\botrule
\end{tabular} \label{param}}
\end{table}
\section{The relativistic effects investigated}
The Einstein pericentre precession is undoubtedly the most famous
relativistic orbital effect, but it is not the only one. Indeed,
also the mean anomaly $\mathcal{M}$ undergoes a secular advance
which is even larger than that experienced by the pericentre. For
small eccentricities it is\cite{Ior05b} \eqi\dot{\mathcal{M}}_{\rm
GE}\sim-\rp{9nGM}{c^2 a\sqrt{1-e^2}},\lb{manom}\eqf  where $G$ is
the Newtonian constant of gravitation, $M$ is the mass of the
central body which acts as source of the gravitational field, $a$,
$e$ and $n=\sqrt{GM/a^3}$ are the semimajor axis, the eccentricity
and the Keplerian mean motion, respectively, of the satellite's
orbit. \Rfr{manom}, which yields a rate of -130 arcseconds per
century for Mercury, has been obtained by using the the standard
isotropic radial coordinate $r$ related to the Schwarzschild
coordinate $r^{'}$ by $r^{'}=r(1+GM/2c^2 r)^2$. The validity of
\rfr{manom} has also been numerically checked by integrating over
200 years the Jet Propulsion Laboratory (JPL) equations of motion
of all the planets of the Solar System, which are written in terms
of just the standard isotropic radial coordinate, with and without
the gravitoelectric $1/c^2$ terms in the dynamical force
models\cite{Est71} in order to single out just the post-Newtonian
gravitoelectric effects (E.M. Standish, private communication,
2004). The obtained precessions fully agree with those obtainable
from \rfr{manom}.

As a consequence, the mean longitude\footnote{Here $\omega$ and
$\Omega$ are the argument of pericentre and the longitude of the
ascending node, respectively.}
$\lambda=\omega+\Omega+\mathcal{M}$, which is one of the
non-singular orbital elements used for orbits with small
eccentricities and inclinations like those of most of the Solar
System major bodies as just the Saturnian satellites (see
Table~\ref{param}), experiences a secular advance
\eqi\dot\lambda_{\rm GE}\sim-\rp{6nGM}{c^2 a}\lb{meanlong}.\eqf It
is twice the pericentre advance. On the other hand, because of
$\mathcal{M}$ in the definition of $\lambda$ the systematic
uncertainty in the Keplerian mean motion $n$ must also be
accounted for if a secular rate must be extracted from the
analysis of such an orbital element.
\section{The possibilities offered by the Saturnian system}
Is it possible to measure the gravitoelectric orbital precessions
of the natural satellites of Saturn, in view of the recent
refinements of their ephemerides obtained by analyzing the first
data from the Cassini spacecraft? In this Section we will address
this problem in detail. To this aim, we must confront the
magnitude of the relativistic effects of interest with the major
systematic errors induced by classical forces having the same
signatures (Section~\ref{evzon}), and with the currently available
measurement accuracies (Section~\ref{accur}).
\subsection{The reduction of the impact of the even zonal
harmonics}\label{evzon}
In regard to the first issue, a major source of systematic bias is
represented by the even zonal harmonic coefficients $J_{\ell},
\ell=2,4,6,...$ of the multipolar expansion of the Saturn's
gravitational potential. Indeed, they induce secular precessions
on $\lambda$ \eqi\dot\lambda_{\rm even\ zonals}=\sum_{\ell\geq
2}\dot\lambda_{.\ell}J_{\ell},\eqf  where the coefficients
$\dot\lambda_{.\ell}$ are expressed in terms of the satellite's
orbital elements $a,e,i$ and the mass and the radius of the
central body. Such classical advances must be accurately modelled
because they are much larger than the relativistic ones. Even the
latest determinations of the Saturnian gravity field from some
Cassini data sets\cite{Jacetal05} show that the currently
available models of the even zonal harmonics do not reach the
required accuracy to allow for a measurement of the Einstein
precessions. This fact is clearly shown by Table~\ref{tablong}.
\begin{table}
\tbl{ Mean longitudes of the major Saturnian satellites: general
relativistic gravitoelectric secular precessions $\dot\lambda_{\rm
GE}$, in \asec, mismodelled classical secular precessions
$\delta\dot\lambda_{J_{2}}, \delta\dot\lambda_{J_{4}}$, in \asec,
due to the uncertainties in the even zonal harmonics $J_2$, $J_4$,
systematic errors $\delta n_{GM}$, in \asec, in the Keplerian mean
motions $n$ due to the uncertainty in Saturn's $GM$. The values
$\delta(GM)=1.2$ km$^3$ s$^{-2}$, $\delta J_2=0.4\times 10^{-6}$,
$\delta J_4=3\times 10^{-6}$  have been used.}
{\begin{tabular}{@{}ccccc@{}} \toprule
 Satellite & $\dot\lambda_{\rm GE}$ & $\delta\dot\lambda_{
J_{2}}$
& $\delta\dot\lambda_{J_{4}}$ & $\delta n_{GM}$\\
\colrule
Mimas & -684.670 & 5926.552 & 5465.353 & 793.461\\
Enceladus & -367.207 & 2478.985 &  1390.913 & 546.018 \\
Thetys & -215.374 & 1173.870 & \hphantom{0}429.231 & 396.440 \\
Dione & -116.003 & \hphantom{0}493.910 &  \hphantom{0}110.229 & 273.492\\
Rhea & \hphantom{0}-50.334 & \hphantom{0}153.451 & \hphantom{00}17.558 & 165.721\\
Titan &  \hphantom{00}-6.152 & \hphantom{000}8.101 & \hphantom{000}0.172 & \hphantom{0}46.950\\
Hyperion & \hphantom{00}-3.679 & \hphantom{000}3.942 & \hphantom{000}0.055 & \hphantom{0}34.487\\
Iapetus & \hphantom{00}-0.424 & \hphantom{000}0.186 & $\mathcal{O}(10^{-4})$ & \hphantom{00}9.437\\
Phoebe &  \hphantom{00}-0.016 & \hphantom{000}0.002 & $\mathcal{O}(10^{-7})$ & \hphantom{00}1.361\\
\botrule
\end{tabular} \label{tablong}}
\end{table}
%
%
Indeed, we can note that for Mimas, Enceladus and Thetys the
mismodelled precessions due to $J_2$ and $J_4$ are larger than the
relativistic rates by about one order of magnitude. For Dione,
Rhea, Titan and Hyperion the bias due to $J_4$ is smaller than the
relativistic signal. For Iapetus and Phoebe the even zonal
harmonics of Saturn do not represent a problem. In addition to the
bias due to the $\{J_{\ell}\}$, for the mean longitude there is
also the systematic error $\delta n_{GM}=\delta(GM)/\sqrt{4GM
a^3}$ induced on the Keplerian mean motion $n$ by the uncertainty
in the Saturn's $GM$ which, as can be seen, is larger than the
relativistic shifts for all the satellites.

A way to overcome the problem of the impact of the even zonals is
represented by the linear combination approach\cite{Ior05b,Ior05c}
summarized in the following. Let us write down the expressions of
the measured residuals\footnote{The Keplerian orbital elements are
not directly observable, so here we are using the words `measured
residuals' in an improper sense. We mean, instead, a set of
solve-for parameters of a suitable least-square solution to be
obtained by contrasting all the available observational data to a
set of dynamical force models in which all the $1/c^2$ terms have
purposely been switched off.} of the mean longitudes
$\delta\dot\lambda_{\rm \small{meas}}$ of $N$ chosen satellites in
terms of the mismodelled Keplerian mean motions, of the classical
precessions due to the mismodelled parts of the even zonal
harmonics, and of the relativistic precessions, assumed as totally
unmodelled features of motion
%
%
%
%
\eqi\delta\dot\lambda^{(j)}_{\rm
\small{meas}}=n_{.GM}\delta(GM)+\sum_{\ell\geq
2}^{2(N-2)}\dot\lambda_{.\ell}^{(j)}\delta
J_{\ell}+\dot\lambda^{(j)}_{\rm GE}\xi,\ j=1,2..N, \lb{syst}\eqf
where $\dot\lambda_{.\ell}$ are the coefficients of the classical
precessions of degree $\ell$ of the mean longitude and \eqi
n_{.GM}=\rp{1}{\sqrt{4GM a^3}}.\eqf
 If we look at \rfr{syst} as a
inhomogeneous  algebraic system of $N$ linear equations in the $N$
unknowns $\{\delta(GM);\delta J_{\ell}; \xi\}$ and solve it for
$\xi$, we get a linear combination of the satellites' mean
longitudes residuals which, by construction, is independent of
$\delta(GM)$ and the first $N-2$ even zonal harmonics, and is
sensitive just to the relativistic signatures and to the
perturbations of the remaining, uncancelled even zonal harmonics.
The cancellation of the low-degree even zonals is important also
because in this way one avoids any possible a priori `imprint' of
the relativistic effects themselves via such spherical harmonics.
Indeed, the Saturnian gravity field models are obtained as
least-square solutions using all the available data from the
spacecraft encounters and the satellites' motions around Saturn,
so that the relativistic effects themselves are in some way
included in the solved-for parameters like $\{J_{\ell}\}$.

A possible combination which cancels out $\delta J_2$, $\delta J_4
$ and $\delta(GM)$ is
\eqi\xi=\rp{\delta\dot\lambda_{\rm \small{meas}}^{\rm
Ph}+k_1\delta\dot\lambda_{\rm \small{meas}}^{\rm
Ia}+k_2\delta\dot\lambda_{\rm \small{meas}}^{\rm
Hy}+k_3\delta\dot\lambda_{\rm \small{meas}}^{\rm
Ti}}{\dot\lambda_{\rm GE}^{\rm Ph}+k_1\dot\lambda_{\rm GE}^{\rm
Ia}+k_2\dot\lambda_{\rm GE}^{\rm Hy}+k_3\dot\lambda_{\rm GE}^{\rm
Ti}},\lb{combi2}\eqf with
\begin{equation}
\left\{
\begin{array}{lll}
k_1=-0.1296,\\\\
k_2=0.0224,\\\\
k_3=-0.0067,\\\\
\dot\lambda_{\rm GE}^{\rm Ph}+k_1\dot\lambda_{\rm GE}^{\rm
Ia}+k_2\dot\lambda_{\rm GE}^{\rm Hy}+k_3\dot\lambda_{\rm GE}^{\rm
Ti}=0.0237 \ '' {\rm cy}^{-1}\lb{coe}
\end{array}
\right.
\end{equation}
As can be argued from Table~\ref{tablong}, the impact of the
remaining uncancelled even zonal harmonics $J_6, J_8$ is
negligible.
\subsection{The present-day orbital accuracy}\label{accur}
In regard to the feasibility of such a measurement, it is useful
to consider the down-track shifts $\Delta\tau\sim a\Delta\lambda$
and the present-day accuracy according to the latest Saturnian
ephemerides SAT240\cite{Jac06} over 81 years. The results are in
Table~\ref{err}.
\begin{table}[ph]
\tbl{Major Saturnian satellites: gravitoelectric down-track
shifts, in km, and present-day orbital accuracy, in km, over a
time span of 81 years. See also on the Internet
http://ssd.jpl.nasa.gov/?sat$\_$ephem. }
{\begin{tabular}{@{}cccc@{}} \toprule
Satellite & $\Delta\tau_{\rm
GE}$ (km) & $\delta(\Delta\tau)$ (km)
\\
\colrule
Mimas & -498 & \hphantom{0}50\\
Enceladus & -343 & \hphantom{0}20\\
Thetys &  -249 & \hphantom{0}50\\
Dione & -171 & \hphantom{0}20\\
Rhea & -104 & \hphantom{0}20\\
Titan & \hphantom{0}-29 & \hphantom{0}10 \\
Hyperion & \hphantom{0}-21 & 100\\
Iapetus & \hphantom{00}-6 & \hphantom{0}20\\
Phoebe &  \hphantom{00}-0.8 & \hphantom{0}50\\
\botrule
\end{tabular} \label{err}}
\end{table}
It can be noted that the situation is presently not favorable just
for the satellites used for the combination of \rfr{combi2}. This
fact reflects on \rfr{combi2} itself: indeed the total uncertainty
calculated by summing in quadrature the errors
$\delta(\Delta\tau)$ of Table~\ref{err} with the coefficients
(\ref{coe}) \eqi\left.\rp{\delta\xi}{\xi}\right|_{\rm
\small{meas}}= \frac{ \sqrt{ [ \delta(\Delta\tau)^{\rm Ph} ]^2 + [
k_1\delta(\Delta\tau)^{\rm Ia} ]^2 + [ k_2\delta(\Delta\tau)^{\rm
Hy} ]^2 + [ k_3\delta(\Delta\tau)^{\rm Ti} ]^2 } }{
\Delta\tau_{\rm GE}^{\rm Ph}+k_1\Delta\tau_{\rm GE}^{\rm
Ia}+k_2\Delta\tau_{\rm GE}^{\rm Hy}+k_3\Delta\tau_{\rm GE}^{\rm
Ti} }\eqf is $\sim 10^3$ times larger than the gravitoelectric
shift over 81 years.

About the use of the inner satellites, for which the relativistic
shifts $\Delta\tau_{\rm GE}$ are larger than the errors
$\delta(\Delta\tau)$, Table~\ref{tablong} shows that it is not
possible to use their mean longitudes without combining them
because both the even zonal mismodelled shifts and the bias of $n$
are quite larger. On the other hand, it turns out that also the
linear combination approach does not yield good results. Indeed,
the root-sum-square of the errors of a combination with Enceladus,
Dione, Rhea and Titan is $10^3$ times larger of the relativistic
combined shifts.
\section{Discussion and conclusions}
In regard to the measurability of the general relativistic
gravitoelectric orbit advances in the Saturn's system of natural
satellites, it turns out that the present-day improvements in
their ephemerides by the Cassini spacecraft are not yet sufficient
to detect such post-Newtonian effects.

As a general rule, in regard to the single satellites, an about
one order of magnitude improvement in the knowledge of the
parameters of the Saturnian gravity field like the even zonal
harmonics $J_{\ell}$ and $GM$ would be required to make the
competing classical bias smaller than the relativistic
precessions. This would make possible to successfully analyze
separately the mean longitudes of each of Mimas, Enceladus,
Thetys, Dione, Rhea and Titan because the present-day down-track
errors are already smaller than the relativistic shifts. The use
of the mean longitudes of Hyperion, Iapetus and Phoebe, for which
the Saturnian even zonals represent a relatively less important
problem, is strictly related to a one-two orders of magnitude
improvements in the down-track parts of their orbits.

The linear combination approach yields, in principle, good results
in reducing the impact of the mismodelling in the even zonals and
$GM$. The main problem with such strategy relies in the down-track
orbital accuracy. Indeed, the down-track errors weighted by their
combination's coefficients must be summed in quadrature, while the
weighted relativistic shifts are combined with their proper own
signs, so that with the present-day precision the overall error
amounts to $10^3$.



\end{document}